\documentclass[aps,pra,twocolumn,letterpaper,10pt,superscriptaddress,floatfix]{revtex4-1}

\usepackage{dsfont}
\usepackage{amsmath}
\usepackage{amsfonts}
\usepackage{amssymb}
\usepackage{bbm}

\usepackage{multirow}

\usepackage{color}
\usepackage{xcolor}

\usepackage{textgreek}

\usepackage{mathrsfs}

\usepackage{fullpage}

\usepackage[utf8]{inputenc}

\usepackage{etex}

\usepackage{mathtools}
\usepackage{xspace}
\usepackage{placeins}
\usepackage{dsfont}
\usepackage{float}

\usepackage{graphicx}

\usepackage{bm}

\newcommand{\R}{\mathbbm{R}}

\newcommand{\br}{{\bm r}}
\newcommand{\bc}{{\bm c}}
\newcommand{\bs}{{\bm s}}
\newcommand{\bn}{{\bm n}}

\newcommand{\add}{a_\mathrm{dd}}

\newcommand{\initial}{\mathrm{i}}
\newcommand{\final}{\mathrm{f}}

\newcommand{\lb}{\mathrm{lb}}
\newcommand{\ub}{\mathrm{ub}}

\newcommand{\ext}{\mathrm{ext}}
\newcommand{\eff}{\mathrm{eff}}

\newcommand\rect{\operatorname{rect}}

\newcommand{\NParticles}{\mathcal{N}}
\newcommand{\NTimeSteps}{N}

\usepackage{amsmath}

\makeindex

\begin{document}

\title{Optimal control of the self-bound dipolar droplet formation process}





\author{J.-F.~Mennemann}
\email{jfmennemann@gmx.de}
\affiliation{Wolfgang Pauli Institute c/o Faculty of Mathematics, University of Vienna, Oskar-Morgenstern-Platz 1, 1090 Vienna, Austria}
\author{T.~Langen}
\affiliation{{5.~Physikalisches Institut} and Center for Integrated Quantum Science and Technology (IQST),
Universit\"at Stuttgart,
Pfaffenwaldring 57, 70550 Stuttgart, Germany}
\author{L.~Exl}
\affiliation{Wolfgang Pauli Institute c/o Faculty of Mathematics, University of Vienna, Oskar-Morgenstern-Platz 1, 1090 Vienna, Austria}
\author{N.~J.~Mauser}
\affiliation{Wolfgang Pauli Institute c/o Faculty of Mathematics, University of Vienna, Oskar-Morgenstern-Platz 1, 1090 Vienna, Austria}

\begin{abstract}

Dipolar Bose-Einstein condensates have recently attracted much attention in the world of quantum many body experiments.
While the theoretical principles behind these experiments are typically supported by numerical simulations,
the application of optimal control algorithms could potentially open up entirely new possibilities.
As a proof of concept,
we demonstrate that the formation process of a single dipolar droplet state
could be dramatically accelerated using advanced concepts of optimal control.
More specifically, our optimization is based on a multilevel B-spline method 
reducing the number of required cost function evaluations 
and hence significantly reducing the numerical effort.
Moreover, our strategy allows to consider box constraints on the control inputs in a concise and efficient way.
To further improve the overall efficiency, we show how to evaluate the dipolar interaction potential
in the generalized Gross-Pitaevskii equation without sacrificing the spectral convergence rate of the
underlying time-splitting spectral method.

\end{abstract}
  	

\maketitle

\section{Introduction}
\label{sec:introduction}

Optimal control has meanwhile become a widely used method 
for the efficient and robust manipulation of quantum systems in various 
applications~\cite{Li_2011, glaser_2015, goerz_2017, machnes_2018} and, 
as such, is expected to play a key role in many emerging quantum technologies 
like sensing or computing~\cite{waldherr_2014, dolde_2014,noebauer_2015}. 
There is also particular interest in the optimization of the intricate processes in complex quantum 
many-body systems~\cite{hohenester_2007,doria_2011}.
In this context, the feasibility of optimal control algorithms 
has been experimentally demonstrated for 
lattice systems and Bose-Einstein condensates~\cite{buecker_2011,vanfrank_2016}. 

The first numerical simulation demonstrating that optimal control provides an efficient way 
to realize the transfer of a Bose-Einstein condensate (BEC) to a desired target state was presented 
in a pioneering work by Hohenester \emph{et al}~\cite{hohenester_2007}.
Mathematical aspects and computational improvements
have been published shortly thereafter in \cite{winckel_2008}, with numerical implementations freely available~\cite{hohenester_2014}. 
While early studies \cite{grond_2009}, \cite{jaeger_2013}, \cite{buecker_2013}, \cite{jaeger_2014} 
focussed on low-dimensional BECs to limit the computational costs, 
optimal control of the full three-dimensional Gross-Pitaevskii equation (GPE) was presented in \cite{mennemann_2015}. 
Moreover, the original approach was generalized to include multiple control inputs. 
In the meantime, various computational and technical aspects, 
as well as different numerical methods have been considered in the literature.
These considerations are not limited to numerical simulations but 
have also been successfully applied in real physical 
experiments~\cite{buecker_2011, vanfrank_2014, vanfrank_2016}.

In the present work we go beyond the isotropic contact interactions typically found in such BECs 
and study the dynamics in BECs with long-range, anisotropic 
dipolar interactions \cite{lahaye_2008,maier_2015,baier_2016,chomaz_2018}. 
Moreover, we also go beyond the 3D mean-field treatment and include the Lee-Huang-Yang 
beyond mean-field correction, which has been shown to lead to the emergence of new and unexpected states of matter, 
including self-bound quantum droplets \cite{baillie_2016,schmitt_2016,chomaz_2016} 
and self-organized striped states \cite{wenzel_2017}.

As a proof of concept, we consider the formation of a self-bound dipolar droplet state.
In particular, we demonstrate that the single droplet formation process studied in~\cite{baillie_2016} 
can be significantly accelerated using efficient numerical methods in combination 
with advanced concepts of optimal control.

Our numerical simulations exploit two novel and effective innovations: 
First, we show how to solve the generalized GPE describing BECs including dipolar 
interactions and the Lee-Huang-Yang mean-field correction.
More precisely, we show how to apply the popular and highly efficient 
time-splitting spectral (split-step Fourier) method
without sacrificing the spectral convergence rate with respect to the 
discretization of the spatial operators.
Second, we turn the underlying infinite dimensional control problem into a 
finite-dimensional nonlinear optimization task using
a B-spline control vector parameterization.
In this context, we also develop a multilevel refinement strategy to enhance 
the speed of convergence of the underlying optimization algorithm.

Optimal control of systems governed by ordinary or partial differential equations
has a long history in mathematics, science and technology
\cite{lions_1971, troeltzsch_2010, gerdts_2011}.
In order to classify and understand the approach proposed in this article it is useful to give 
a short review on the two predominant methods frequently found in the literature 
of quantum optimal control, and in particular, in the field of BEC experiments. 

The first approach is based on a reduced cost functional 
\begin{equation}
\label{eq:reduced_cost_functional}
\hat{\mathcal{J}}_1(u) = \mathcal{J}_1(\psi_u, u),
\end{equation}
where $\mathcal{J}_1(\psi, u)$ is a  suitably selected cost functional describing the optimal control problem
and $u:[0,T]\rightarrow \mathbbm{R}$ denotes a given control input.
Furthermore, $\psi_u$ denotes the time-evolution of the wave function corresponding
to a solution of the state equation using $u$ as the control input.
In the context of BEC experiments, 
the state equation would be given by the GPE.
The state equation is considered to be a constraint to the optimization problem.
Using the calculus of variations an additional adjoint
equation is derived which allows for an efficient computation of the gradient 
of~$\hat{\mathcal{J}}_1$ needed to improve the given control input $u$.
Until then, no numerical approximations are involved.
Finally, the state and adjoint equation are solved using suitable numerical discretizations.
From a mathematical point of view, this approach appears to be very elegant,
however, the derivation and discretization of an additional adjoint equation
is a highly non-trivial task and must be repeated whenever the state equation or the cost functional is modified.
Moreover, one should bear in mind that the numerical solution of the adjoint equation is typically more 
expensive than the solution of the state equation, see, e.g., \cite{mennemann_2015}.

In the second approach, the reduced cost functional~\eqref{eq:reduced_cost_functional}
is replaced by a reduced cost function
\begin{equation}
\label{eq:reduced_cost_function}
\hat{\mathcal{J}}_2(\bc) = \mathcal{J}_2(\psi_{\bc}, \bc)
\end{equation}
with $\bm{c} = [c_1, \ldots,c_K]$ being a vector collecting the $K$ coefficients 
of a suitable control vector parameterization~(CVP).
As an example, we consider a simple sum of sines parameterization
\begin{equation}
\label{eq:sum_of_sines}
u(\bm{c}, t) = u_0 + (u_T- u_0) t/T + \sum_{k=1}^K c_k \sin(k \pi t / T),
\end{equation}
where 
\begin{equation}
\label{eq:initial_final_condition}
u_0=u(t=0), \quad u_T=u(t=T)
\end{equation}
denote the initial and the final conditions 
of the control input, respectively. 
The first argument of $\mathcal{J}_2(\psi_{\bm{c}}, \bm{c})$ denotes the time evolution of 
the wave function corresponding to a solution of the GPE using $u(\bm{c},t)$
as the control input.
It should be noted that $\hat{\mathcal{J}}_1$ in~\eqref{eq:reduced_cost_functional}
is a functional, whereas $\hat{\mathcal{J}}_2$ in~\eqref{eq:reduced_cost_function} is a function. 
Therefore, the previously infinite-dimensional optimal control problem has been mapped to 
a finite-dimensional nonlinear Program (NLP) which can be solved using various numerical 
methods like the popular Nelder-Mead algorithm~\cite{nelder_mead_1965}.

With these preconditions CVP methods are easy to implement as the only requirement is
a solver for the forward problem, i.e., a numerical method to solve the given 
differential or partial differential equation.
Not surprisingly, CVP methods are frequently applied in classical optimal control applications 
and have been used long before they were applied 
to quantum optimal control problems, see, e.g., \cite{kraft_1985, schlegel_2005, hadiyanto_2008}.

A CVP method based on a randomized Fourier basis
is proposed as a general and versatile quantum optimal control technique in~\cite{caneva_2011, rach_2015}.
However, like most randomized algorithms, the proposed method converges extremely slowly
as was demonstrated recently in an extensive benchmark problem \cite{sorensen_2018}.
In fact, the simple ansatz in \eqref{eq:sum_of_sines} seems to be more efficient
and could be easily applied in a multistart optimization procedure to explore the space 
of possible solutions in a more rigorous way.

Often, a surprisingly small number of coefficients $K$ is sufficient to obtain a
high-quality solution of the underlying control problem. 
However, in some quantum control applications~\cite{sorensen_2018} the number of 
coefficients needed to approximate pulses of very small duration can easily exceed 
$K \approx 20$.
For such large values of $K$, derivative-free methods like the Nelder-Mead algorithm
are known to be less efficient.

In principle, the gradient of the reduced cost function~\eqref{eq:reduced_cost_function}
with respect to the coefficients $\bm{c}$
can always be approximated using simple finite difference formulas.
To this end, the forward problem needs to be solved at least
$K+1$ times.
More efficient methods to compute the gradient are presented in~\cite{winckel_2008}
and \cite{sorensen_2018} for a CVP based on Chebyshev polynomials and a sum of sines
ansatz, respectively. 
Generally speaking, the approach is not limited to a specific CVP, however, 
it also requires the derivation and discretization of an additional adjoint equation.
Once a numerical method to evaluate the gradient is available, the NLP can be solved
using most effective implementations of quasi-Newton algorithms (BFGS) or
sequential quadratic programming (SQP) methods \cite{nocedal_wright_2006}.  

While the CVP defined in Eqn.~\eqref{eq:sum_of_sines} works well
in the context of unconstrained optimization problems, it results in suboptimal
solutions if the control input is subject to constraints
\begin{equation}
\label{eq:lower_upper_bounds}
u_\mathrm{lb} \leq u(\bc, t) \leq u_\mathrm{ub}, \quad t \in [0,T]
\end{equation}
naturally present in realistic control applications.
This applies in particular if the control horizon $T$ is chosen to be small.
The reason for this is that, as the $T$ becomes smaller, the corresponding
optimal control tends to switch from one extreme to the other.
In minimum-time problems the corresponding control is referred to as a bang-bang 
solution.
In our application, we are not aiming at solving a minimum-time optimal control problem.
However, we will choose the time horizon $T$ small enough that the effect of the
constraints~\eqref{eq:lower_upper_bounds} starts to play an important role.
More precisely, we will see that the control inputs resemble bang-bang solutions, i.e.,
the lower and upper bounds become active for time intervals of non-vanishing lengths.
We would like to add that these kind of controls cannot be obtained using
the simple ansatz given in Eqn.~\eqref{eq:sum_of_sines} as this would require an infinite 
number of sine waves $K$.

As an alternative, we propose to parameterize the control input by 
a superposition of B-spline basis functions~\cite{d_boor_1978}.
As a result of their compact support, 
B-spline basis functions represent a much more favorable ansatz with respect 
to Eqn.~\eqref{eq:lower_upper_bounds}.
Moreover, Dirichlet boundary conditions~\eqref{eq:initial_final_condition} can be implemented easily 
and B-spline basis functions are ideally suited 
for multiresolution function approximations.
In fact, we will demonstrate that a multilevel refinement strategy can significantly improve the 
convergence of the overall optimization procedure.

The paper is organized as follows:
Section~\ref{sec:generalized_GPE} covers the generalized GPE, 
the phase diagram of a trapped dipolar Bose gas and all relevant physical parameters.
In the first part of Section~\ref{sec:numerical_discretization}, we review the 
time-splitting spectral method (TSSM) frequently used to solve the ordinary GPE.
Moreover, we introduce the notation concerning the spatial and temporal discretizations
and explain why the TSSM can be easily applied to the generalized GPE as well.
The second part of Section~\ref{sec:numerical_discretization} is devoted 
to spectrally accurate numerical methods for the evaluation of the dipolar interaction potential.
In Section~\ref{sec:multilevel_B_spline_method}, we introduce the multilevel B-spline method
which is finally applied to the dipolar droplet formation process in Section~\ref{sec:application}.

\section{Generalized Gross-Pitaevskii Equation}
\label{sec:generalized_GPE}

Our considerations are based on the description of dipolar Bose-Einstein condensates given by 
the generalized Gross-Pitaevskii equation \cite{baillie_2016}
\begin{equation}
\label{eq:generalized_gpe}
\begin{aligned}
&\imath \hbar \frac{\partial \psi(\br, t)}{\partial t}
=
\bigg[
-\frac{\hbar^2 \nabla^2}{2 m}
+ V_\mathrm{ext}(\br, t)
+ g(t) |\psi(\br, t)|^2
\\
&+ \Phi(\br, t) 
+ \gamma_\mathrm{qf}(t) |\psi(\br, t)|^3
- \frac{\imath \hbar L_3}{2} |\psi(\br, t)|^4
\bigg] \psi(\br, t),
\end{aligned}
\end{equation}
where $\psi$ is the condensate wave function with 
initial state $\psi_0 = \psi(t=0)$.
The density of the initial state is assumed to be normalized to the
number of atoms
$\NParticles_0 = \| \psi_0 \|^2$.
Furthermore, the contact interaction strength is given by
$g = 4 \pi a_s \hbar^2 / m$, where $a_s$ denotes the s-wave scattering length and $m$ the mass of the atoms.
In the numerical simulations below, the external potential
\begin{equation}
\label{eq:V_ext}
V_\mathrm{ext}(\br, t) = \frac{1}{2} m \omega_{\rho}^2(t) (x^2 + y^2) + \frac{1}{2} m \omega_z^2(t) z^2
\end{equation}
is a function of the radial and longitudinal trapping frequencies $\omega_\rho$ and $\omega_z$, which 
can vary with time and are considered to be freely adjustable
parameters.
Dipole-dipole interactions are taken into account via the dipolar interaction potential
\begin{subequations}
\label{eq:dipolar_interaction}
\begin{equation}
\label{eq:Phi}
\Phi(\br, t) = \frac{\mu_0 \mu^2}{4 \pi} \int_{\mathbbm{R}^3} U(\br-\br') |\psi(\br',t)|^2 \, d\br',
\end{equation}
where $\mu$ is the magnetic moment of the considered atoms, $\mu_0$ is the vacuum permeability and
\begin{equation}
\label{eq:kernel_U}
U(\br)
=
\frac{1 - 3 \cos^2 \theta}{r^3},
\end{equation}
\end{subequations}
with $r = (x^2 + y^2 + z^2)^{1/2}$.
Moreover, $\theta$ denotes the angle between $\br$ and the polarization axis 
$\bn = (n_1, n_2, n_3)^\top$
which is assumed to be normalized, i.e.,
$(n_1^2 + n_2^2 + n_3^3)^{1/2} = 1$.
Within a local density approximation \cite{lima_2011}
the correction term $\gamma_\mathrm{qf} |\psi|^3$
is used to include the effect of quantum fluctuations \cite{lima_2011, waechtler_2016, schmitt_2016, baillie_2016}.
The prefactor
\begin{equation}
\label{eq:gamma_qf}
\gamma_\mathrm{qf} = \frac{32}{3} g \sqrt{\frac{a_s^3}{\pi}} \bigg( 1 + \frac{3}{2} \frac{\add^2}{a_s^2} \bigg)
\end{equation}
depends on the 
dipolar length $a_\mathrm{dd} = m \mu_0 \mu^2 / (12 \pi \hbar^2)$ \cite{lahaye_2009}.
Finally, three-body loss processes are taken into account by the term
$-(\imath \hbar L_3 / 2) |\psi|^4$ with $L_3$ being the loss coefficient.

\begin{figure}
\centering
\includegraphics[width=0.45\textwidth]{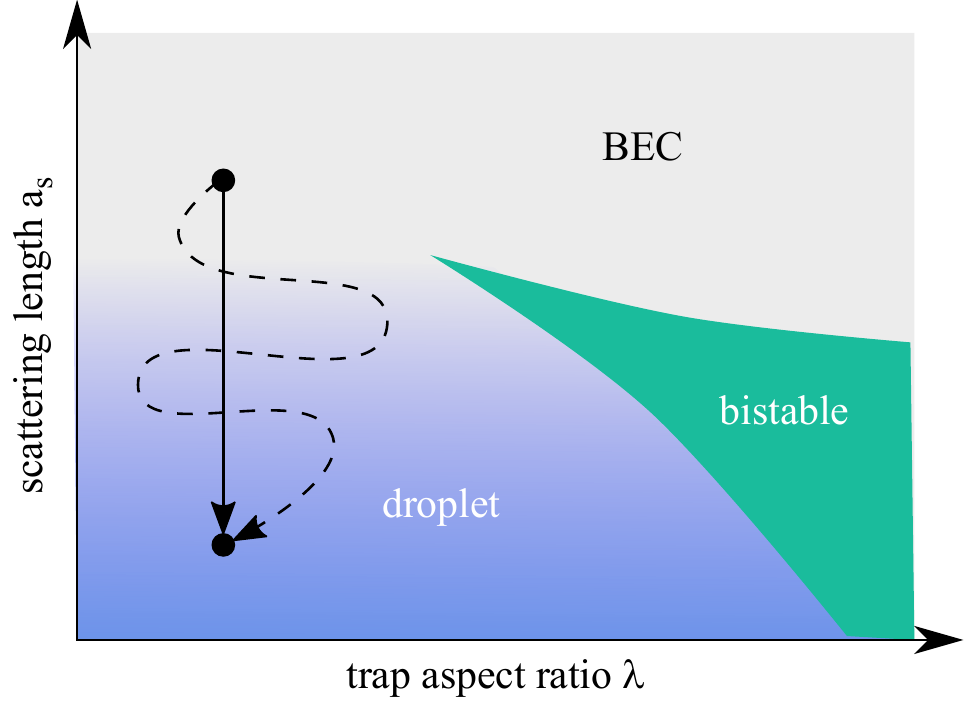}
\caption{
Phase diagram of a trapped dipolar Bose gas as a function of scattering length $a_s$ and trap aspect ratio $\lambda$  \cite{baillie_2016,waechtler_2016, schmitt_2016}. 
The system is initially a Bose-Einstein condensate. A linear decrease of $a_s$ (solid arrow) smoothly transforms this condensate into a stable dipolar droplet. 
However, to avoid excitations of the droplet this trajectory needs to be followed adiabatically, 
resulting in significant three-body atom loss. 
In this work we compute much faster optimal trajectories (indicated by the dashed arrow), 
which use both $a_s$ and $\lambda$ as control parameters, while avoiding a bistable region 
in the phase diagram where both BEC and droplet solutions coexist.
}
\label{fig:phasediagram}
\end{figure}

The validity of this approximation has previously been checked using Monte-Carlo simulations~\cite{saito_2016}. 
The effect of quantum fluctuations is usually very small, however it can be dramatically enhanced when the 
mean-field contributions of contact and dipolar interactions nearly cancel. 
This is the case, for example, during the collapse following a motional instability of the system~\cite{waechtler_2016}.
 
The resulting phase diagram for the system is schematically shown in Fig.~\ref{fig:phasediagram}~\cite{baillie_2016}. 
For high scattering length $a_s$ the system is a dipolar BEC, for low $a_s$ it forms a single stable quantum droplet. 
Above a critical aspect ratio $\lambda_c\sim 1.87$ the system is bistable~\cite{ferrier_2018}. Lowering $a_s$ into this bistable region triggers a motional instability that leads to the formation of arrays containing several droplets. 

Our aim is to optimize the self-bound dipolar droplet formation process presented in~\cite{baillie_2016}. 
This work proposed to linearly decrease the scattering length, 
which - for appropriate trap aspect ratios - connects the BEC and the self-bound droplet by a smooth crossover. 
This is followed by a linear decrease of the trap frequencies to turn off the trapping potential 
and reveal the self-bound nature of the droplet. 
However, adiabatically following such a linear trajectory requires a long timescale and thus 
leads to non-negligible atom loss. 
Alternatively, fast linear ramps typically lead to unwanted excitations. 
As we will show in the following, optimal control can help with both these problems, 
preparing droplets without observable excitations on very short timescales. 

Like in \cite{baillie_2016} we consider $^{164}$Dy atoms. 
We therefore choose
$m = 163.93$\,u, $\mu = 9.93\,\mu_\mathrm{B}$ and $L_3 = 1.2 \times 10^{-41}\,\mathrm{m}^6 s^{-1}$, 
with $u$ the atomic mass unit. 
Initially, 
$\NParticles_0 = 10^4$ atoms are trapped in the harmonic potential at
$t=0$.
Furthermore, $a_s, \omega_\rho$ and $\omega_z$ are considered to be control inputs 
which are subject to the initial and final conditions
\begin{subequations}
\label{eq:initial_final_values}
\begin{align}
a_s(t=0) &= a_{s, \initial},
&
a_s(t=T) &= a_{s, \final}, \\
\omega_\rho(t=0) &= \omega_{\rho, \initial},
&
\omega_\rho(t=T) &= \omega_{\rho, \final}, \\
\omega_z(t=0) &= \omega_{z, \initial},
&
\omega_z(t=T) &= \omega_{z, \final}
\end{align}
\end{subequations}
as well as the constraints
\begin{subequations}
\label{eq:limits_control_inputs}
\begin{align}
a_s^\lb &\leq a_s(t) \leq a_s^\ub,
\\
\omega_\rho^\lb &\leq \omega_\rho(t) \leq \omega_\rho^\ub,
\\
\omega_z^\lb &\leq \omega_z(t) \leq \omega_z^\ub
\end{align}
\end{subequations}
for $t \in [0, T]$ where $T$ denotes the final time of the optimal control problem considered below. 
All remaining parameters of
Eqn.~\eqref{eq:initial_final_values} and Eqn.~\eqref{eq:limits_control_inputs} are summarized as follows:
\begin{align*}
a_{s,\initial} &= 130\,a_0, 
& 
a_{s,\final} &= 80\,a_0, \\
a_s^\lb      &= 80\,a_0, 
&
a_s^\ub      &= 130\,a_0, \\
\omega_{\rho, \initial} &= 2 \pi \times 70\,\textrm{Hz},
& 
\omega_{\rho, \final} &= 2 \pi \times 0\,\textrm{Hz}, \\
\omega_\rho^\lb       &= 2 \pi \times 0\,\textrm{Hz},
&
\omega_\rho^\ub       &= 2 \pi \times 318.3\, \mathrm{Hz},\\
\omega_{z,\initial}   &= 2 \pi \times 52.5\, \textrm{Hz},
&
\omega_{z,\final}     &= 2 \pi \times 0\,\textrm{Hz}, \\
\omega_z^\lb          &= 2 \pi \times 0\,\textrm{Hz},
&
\omega_z^\ub          &= 2 \pi \times 318.3\, \mathrm{Hz}.
\end{align*}

In experiments $a_s, \omega_\rho$ and $\omega_z$
can be changed using a Feshbach resonance and the laser intensities 
of the trapping lasers, respectively. 
The initial value of $a_s$ in \cite{baillie_2016} is given by $a_{s,\initial}=130\,a_0$, 
which is linearly decreased to $a_{s,\final} = 80\,a_0$. 
In our optimization we restrict ourselves to this range of scattering lengths to take into account 
the fact that Feshbach resonances in lanthanide atoms are typically very narrow and their 
distribution highly complex~\cite{maier_2015}. 
In contrast to alkali atoms the s-wave scattering length can thus only be tuned over a 
limited range, which has to be taken into account in the control sequence.

Moreover, we also choose the same initial configuration as in \cite{baillie_2016} 
for the external confinement potential at time $t=0$,
i.e., $\omega_{\rho, \initial} = 2 \pi \times 70$\,Hz 
and $\omega_{z,\initial}=\lambda\, \omega_{\rho, \initial}$ with $\lambda = 0.75$ being the trap aspect ratio. 
As for the scattering length we include limits for these trapping frequencies, 
which are bounded by the available laser power in experiments.
Since we are aiming at creating a self-bound droplet state, the external potential is required to vanish at
$t=T$ and thus $\omega_{\rho, \final} = \omega_{z, \final} = 0$. 
In all our calculations, the polarization axis of the dipoles is aligned with the
positive $z$-direction, i.e., $\bn = (0, 0, 1)^\top$.

\section{Numerical Discretization}
\label{sec:numerical_discretization}

\subsection{Time-Splitting Spectral Method}

Our strategy to solve the optimal control problem considered in Section~\ref{sec:application}
requires hundreds of simulations of the generalized GPE~\eqref{eq:generalized_gpe}.
A variety of numerical methods is described in the literature to compute the time-evolution
of nonlinear Schr\"odinger equations.
The most straightforward way is to first discretize the spatial operators using
finite difference approximations and then to apply explicit Runge-Kutta methods 
(like the classical Runge-Kutta method of order four) to integrate the resulting system of
ordinary differential equations in time.
While this approach is simple to implement, very small time step sizes are needed to
ensure numerical stability \cite{caplan_2013}.
Much larger time step sizes can be used if the time-integration is based on
implicit methods.
A proven method in the context of quantum dynamical simulations is the Crank-Nicolson scheme, 
see e.g. \cite{bao_numerical_solution_2003}.
This latter approach represents an excellent choice in one spatial dimension, however, in two or
three spatial dimensions significant computational resources are needed
to solve the corresponding linear systems of equations.
Finally, an efficient and frequently used method to compute the time-evolution of the 
classical GPE is the time-splitting spectral (TSSM) 
method \cite{bao_numerical_solution_2003, bao_numerical_study_2003, thalhammer_high-order_2009}.
We recall that the most frequently used implementation of the TSSM is based on the 
Strang time splitting~\cite{bao_numerical_solution_2003}, i.e.,
\begin{subequations}
\begin{equation}
\label{eq:strang_splitting}
\psi(t_{n+1}) 
\approx 
e^{- \imath B^+ \triangle t / 2} e^{-\imath A \triangle t} e^{-\imath B^- \triangle t / 2} \psi(t_n)
\end{equation}
with the operators
\begin{equation}
\label{eq:operators_A_B}
A = - \hbar \nabla^2/(2 m), \quad B^\pm = V_\eff / \hbar
\end{equation}
and the effective potential energy
\begin{equation}
\label{eq:eff_potential_gpe}
V_\eff = V_\ext + g |\psi^\pm_n|^2.
\end{equation}
\end{subequations}
Moreover, $t_n = n \triangle t$ 
with $\triangle t$ being the time step size and
$n = 0,...,\NTimeSteps$ such that\ $N \triangle t = T$.
According to \eqref{eq:strang_splitting}, the $n$th time step consists of three sub-steps.
First, we solve
\begin{equation}
\label{eq:tssm_potential_part}
\imath \hbar \partial_t\psi = (V_\ext + g |\psi(t_n)|^2) \psi
\end{equation}
for a duration of $\triangle t / 2$ using $\psi_n^- = \psi(t_n)$ as initial value.
The result is used as initial value for the free Schr\"odinger equation 
\begin{equation}
\label{eq:tssm_free_schroedinger}
\imath \hbar \partial_t\psi = - \hbar^2 \nabla^2/(2 m) \psi,
\end{equation}
which is then solved for a duration of $\triangle t$;
the result is $\psi_n^+$.
Finally, $\imath \hbar \partial_t \psi = (V_\ext + g |\psi_n^+|^2) \psi$ is solved again 
with initial value $\psi_n^+$, again for a duration of $\triangle t / 2$.
The result of the third sub-step is taken as approximation of $\psi(t_{n+1})$.

The spatial discretization is based on the grid points
$x_j = -L_x/2 + j_x \triangle x$,  $y_j = -L_y/2 + j_y \triangle y$
and $z_j = -L_z/2 + j_z \triangle z$
with
$\triangle x = L_x/J_x$, $\triangle y = L_y/J_y$ and $\triangle z = L_z/J_z$.
The parameters $L_x$, $L_y$ and $L_z$ are chosen sufficiently large such that
the wavefunction is compactly supported within the computational domain
\begin{equation}
\label{eq:Omega}
\Omega = [-L_x/2, L_x/2] \times [-L_y/2, L_y/2] \times [-L_z/2, L_z/2].
\end{equation}
Moreover, $j_x = 0,...,J_x-1$, $j_y = 0,...,J_y-1$, $j_z = 0,...,J_z-1$ 
and $J = J_x J_y J_z$ denotes the total number of spatial grid points.

In a numerical implementation the solution of \eqref{eq:tssm_potential_part} 
requires a pointwise evaluation of the exponential function on a real valued array as well as
a pointwise multiplication of a real and a complex valued array.
The free Schr\"odinger equation \eqref{eq:tssm_free_schroedinger} can easily be solved in Fourier space
and the numerical implementation requires the application of two
discrete Fourier transforms (forward and inverse) and
a pointwise multiplication of two complex valued arrays, see e.g. \cite{bao_numerical_solution_2003}.
For the sake of numerical efficiency, 
the discrete Fourier transforms are computed using the Fast Fourier Transform (FFT)
and hence the numerical costs of a single time step are of order $\mathcal{O}(J \log J)$.

The TSSM outlined above (Strang splitting) is of second order in time.
Moreover, the spatial derivatives in the free Schr\"odinger equation are discretized with
spectral accuracy, meaning that the numerical error corresponding to the discretization of the 
Laplacian operator 
decreases at an exponential rate with respect to the number of spatial grid points.
It should be noted that these convergence rates are based on the assumption that
the initial wave function and the external confinement potential are sufficiently smooth.
However, in most experimentally relevant simulation scenarios these assumptions are satisfied 
and thus the number of spatial grid points can be reduced drastically in comparison with
finite difference approximations.
Moreover, the TSSM allows for the application of large time step sizes $\triangle t$ and as a result
the total number of time steps needed to compute the time evolution of the GPE
is significantly lower than in the case of the above-mentioned explicit time-integration methods.
To summarize, the TSSM represents a very efficient numerical method for the solution of the GPE.

In the following, we show that the TSSM can also easily be applied to 
the generalized GPE in~\eqref{eq:generalized_gpe}. 
To this end, we only need to substitute the effective potential 
energy in \eqref{eq:eff_potential_gpe} by
\begin{equation}
\label{eq:V_eff_1st}
V_\eff = V_\ext + g |\psi^\pm_n|^2 + \Phi_n^\pm + \gamma_\mathrm{qf} |\psi_n^\pm|^3 
- \frac{\imath \hbar L_3}{2} |\psi_n^\pm|^4
\end{equation}
with $\Phi_n^\pm$ being defined in analogy to \eqref{eq:Phi} using $\psi_n^\pm$ instead of $\psi$.
The terms $\gamma_\mathrm{qf} |\psi^\pm|^3$ and $-i \hbar L_3 / 2 \, |\psi^\pm|^4$
can be included at almost no additional numerical costs.
However, the evaluation of $\Phi$ in \eqref{eq:Phi} is a non-trivial task
if we want to conserve the spectral convergence rate of the TSSM 
with respect to the discretization of the spatial operators.

\subsection{Evaluation of the Dipolar Interaction Potential}

A simple way to evaluate the dipolar interaction potential $\Phi$ is based on the convolution theorem
which yields
\[
\Phi = \big( \mu_0 \mu^2 / (4 \pi) \big) \mathcal{F}^{-1} \Big[ \mathcal{F}[U] \, \mathcal{F}[\rho] \Big],
\]
where $\mathcal{F}$ denotes the Fourier transform and $\rho$ is the density $\rho=|\psi|^2$.
In a numerical implementation $U$ and $\rho$ are replaced by their discrete 
representations corresponding to the numerical grid of the computational domain $\Omega$.
Moreover, all Fourier transforms are implemented using FFTs in combination with
a zero-padding strategy.
Due to the singularity of $U(\br)$ at $\br = (0,0,0)^\top$ 
the accuracy of the approximation is of second order
with respect to the number of spatial grid points $J$.

To find a discretization of the dipolar interaction potential that converges 
at an exponential rate, 
we first apply the same transformation that has been employed in \cite{bao_efficient_2010}. 
More precisely, we make use of the fact that \eqref{eq:kernel_U} 
can be written~(in the sense of distributions) as 
\begin{equation} 
\begin{aligned}
\label{eq:U_modified}
U(\br) 
&=
-\frac{4 \pi}{3} \delta(\br) - \partial_{\bn \bn} \bigg(\frac{1}{|\br|} \bigg),
\end{aligned}
\end{equation}
where
$
\partial_{\bn} = \bn \cdot \nabla = n_1 \partial_x + n_2 \partial_y + n_3 \partial_z
$
and
$\partial_{\bm{nn}} = \partial_{\bn} \partial_{\bn}$.
Plugging \eqref{eq:U_modified} into \eqref{eq:Phi} yields
\begin{equation*}
\begin{aligned}
\Phi(\br, t) 
&= 
-\frac{\mu_0 \mu^2}{3} |\psi(\br, t)|^2 \\
&\quad - \frac{\mu_0 \mu^2}{4 \pi} \int_{\mathbbm{R}^3} \partial_{\bn\bn} \bigg( \frac{1}{|\br - \br'|} \bigg) 
|\psi(\br',t)|^2
\, d\br'
\end{aligned}
\end{equation*}
or
\[
\Phi(\br, t) 
= 
-g_\mathrm{dd} |\psi(\br, t)|^2
- 3 g_\mathrm{dd}  \partial_{\bn\bn} \varphi(\br, t),
\]
where
$g_\mathrm{dd} = 4 \pi \hbar^2 a_\mathrm{dd} / m = \mu_0 \mu^2 / 3$
and
\begin{subequations}
\label{eq:convolution_poisson}
\begin{equation}
\label{eq:convolution_poisson_1}
\varphi(\br, t) = \int_{\R^3} u(\br-\br')  |\psi(\br', t)|^2 \,d\br'
\end{equation}
with the new kernel
\begin{equation}
\label{eq:convolution_poisson_2}
u(\br)
=
\frac{1}{4 \pi |\br|}.
\end{equation}
\end{subequations}
The normal derivative operator $\partial_{\bn\bn}$ can be evaluated using spectral differentiation
(at the cost of two additional FFTs of size $J=J_x J_y J_z$).
Consequently, the only remaining difficulty is the computation of the 
convolution in~\eqref{eq:convolution_poisson}, which however is a ubiquitous problem 
in computational physics and scientific computing in general.
In fact, the solution of \eqref{eq:convolution_poisson} is equivalent 
to the solution of the free-space Poisson equation
\begin{equation}
\label{eq:poisson_equation}
\Delta \varphi(\br,t) 
= 
- |\psi(\br, t)|^2,
\quad 
\lim_{|\br| \rightarrow \infty} \varphi(\br,t) = 0.
\end{equation}

A variety of efficient numerical methods has been developed to solve the Poisson equation 
in unbounded 
domains~\cite{ethridge_new_2001, aguilar_2005, langstron_free-space_2011, hejlesen_high_2013, 
malhotra_pvfmm_2015, exl_2016, vico_2016, exl_2017}.
While most of them are based on formulation \eqref{eq:convolution_poisson}
it is also possible to employ \eqref{eq:poisson_equation} directly.
This strategy is used by the authors of \cite{bao_efficient_2010} in the context of numerical simulations of dipolar Bose-Einstein condensates.
By replacing the far-field boundary condition in \eqref{eq:poisson_equation}
with a homogeneous Dirichlet boundary condition on the boundary of the computational domain,
the Poisson equation can be solved easily using a sine pseudospectral method.
The method avoids the application of zero-padding.
However, since the potential $\varphi$ in \eqref{eq:poisson_equation} is known to decay slowly,
an extremely large computational domain is needed to keep the modeling error small.

A highly efficient numerical method for the evaluation of nonlocal potentials, including the 3D
Coulomb potential \eqref{eq:convolution_poisson}, has been presented recently~\cite{exl_2016, exl_2017}.
The method is based on a Gaussian-sum approximation of the singular convolution kernel
(combined with a Taylor expansion of the density) 
and it has been shown that for sufficiently smooth right hand sides (densities)
the numerical error  decreases at an exponential rate with respect to the number of 
spatial grid points $J$.
Apart from a relatively expensive precomputation, 
the most time-consuming part of the method is the application of several FFTs 
in combination with a zero-padding strategy.

Another spectrally accurate algorithm was presented almost simultaneously in \cite{vico_2016},
motivated by the truncation technique of \cite{fuesti_molnar_2002, ronen_2006}.
The implementation of the latter method is much simpler but involves a very memory-intensive
precomputation.
However, in our application these limitations are not an issue which is why we employ the second algorithm.
The second algorithm is based on the observation that the solution of \eqref{eq:convolution_poisson} 
is indistinguishable from the solution of
\begin{subequations}
\label{eq:convolution_truncated}
\begin{equation}
\label{eq:convolution_truncated_1}
\varphi(\br, t) = \int_{\R^3} u_L(\br-\br') |\psi(\br', t)|^2 \,d\br'
\end{equation}
where
\begin{equation}
\label{eq:convolution_truncated_2}
u_L(\br) = u(\br) \rect \bigg( \frac{|\br|}{2 L} \bigg)
\end{equation}
\end{subequations}
with the characteristic function
\begin{equation*}
\rect(x)
=
\begin{cases}
1 \quad \mathrm{for} \quad |x| < 1/2 \\
0 \quad \mathrm{for} \quad |x| > 1/2
\end{cases}
\end{equation*}
and $L = \sqrt{L_x^2 + L_y^2 + L_z^2}$.
Since $u_L$ is a compactly supported function the Paley-Wiener theorem implies that its Fourier transform 
\begin{equation*}
\label{u_L_hat}
\hat{u}_L(\bs) = \mathcal{F}\big[ u_L \big](\bs) = 2 \Big( \frac{\sin(L |\bs| / 2)}{|\bs|} \Big)^2
\end{equation*}
is an entire function \cite{vico_2016} (and thus $C^\infty$).
Application of the convolution theorem to \eqref{eq:convolution_truncated} yields
\[
\varphi = \mathcal{F}^{-1} \Big[ \mathcal{F}[u_L] \, \mathcal{F}[\rho] \Big]
\]
which is the starting point for the numerical discretization.
However, since $\hat{u}_L$ is a highly oscillatory function, the complexity of the algorithm significantly increases.
More precisely, $\hat{u}_L$ needs to be sampled using a four-fold sampling rate, which finally requires
the application of FFTs of size $4 J_x 4 J_y 4 J_z$.
Fortunately, these complications only concern the precomputation phase.
During normal operation the algorithm requires the application of two FFTs (forward and inverse) of size 
$2 J_x 2 J_y 2 J_z$ and hence the algorithm is of the same complexity as the simple strategy presented 
at the very beginning of this section.

All considerations in \cite{vico_2016} are based on the assumption that $\Omega$ is a unit box 
with an aspect ratio
\[
\zeta = \max(L_x, L_y, L_z) / \min(L_x, L_y, L_z)
\]
equal to one.
In the application considered below the aspect ratio of the computational domain is given by $\zeta = 2$.
By means of numerical experiments using gaussian density distributions and their known analytical solutions
we found that the algorithm still converges exponentially fast 
if the aspect ratio remains smaller than approximately $\zeta \lessapprox 2.75$.
Furthermore, we found that for even larger aspect ratios, 
the sampling rate in the precomputation phase needs to be increased (six-fold sampling rate for 
$2.75 \lessapprox \zeta \lessapprox 4.5$).


\section{Multilevel B-spline method}
\label{sec:multilevel_B_spline_method}

For the reasons stated in the introduction, we propose to parameterize the control inputs
using B-spline functions
\begin{equation}
\label{eq:cvp_B_spline}
u(\bc,t) = \sum_{k=1}^K c_k N_{k,p}(t),
\end{equation}
where $B_{k,p}$ denotes the $k$th B-spline basis function of polynomial order $p$.
Given a nondecreasing knot vector $\Xi = \{ \xi_1, \xi_2, \ldots, \xi_{K+p+1} \}$
the $K$ basis functions are defined recursively by means of the
Cox-de Boor recursion formula~\cite{d_boor_1978, cottrell_2009}
\begin{align*}
N_{k,p}(\xi)
&=
\frac{\xi-\xi_k}{\xi_{k+p}-\xi_k} N_{k,p-1}(\xi) \\
&\quad \quad \quad + \frac{\xi_{k+p+1}-\xi}{\xi_{k+p+1}-\xi_{k+1}} N_{k+1, p-1}(\xi)
\end{align*}
with
\[
N_{k,0}(\xi)
=
\begin{cases}
1, \quad \textrm{if } \xi_k \leq \xi < \xi_{k+1}, \\
0, \quad \textrm{otherwise}
\end{cases}
\]
being piecewise constant functions.

In the simulations below, we employ cubic B-spline functions, i.e., we set $p=3$.
Moreover, our optimization algorithm is based on the uniform knot vectors
\begin{subequations}
\label{eq:knot_vectors}
\begin{align}
\label{eq:knot_vector_1}
\Xi_1  &= \{0, 0, 0, 0, T,   T, T, T \}, \\
\label{eq:knot_vector_2}
\Xi_2  &= \{0, 0, 0, 0, T/2, T, T, T, T \}, \\
\label{eq:knot_vector_3}
\Xi_3 &= \{0, 0, 0, 0, T/4, 2T/4, 3T/4, T, T, T, T \}
\end{align}
and
\begin{equation}
\label{eq:knot_vector_4}
\begin{aligned}
\Xi_4 &= \{0, 0, 0, 0, T/8, 2T/8, 3T/8, 4T/8, \\ 
&\qquad 5T/8, 6T/8, 7T/8, T, T, T, T \}
\end{aligned}
\end{equation}
\end{subequations}
corresponding to four different levels of refinement.
These knot vectors are said to be open since the first and last knot values appear $p+1$ times.
Most applications of B-spline basis functions are based on open knot vectors.
This is due to the fact that basis functions formed from open knot vectors are
interpolatory at the ends of the interval $[\xi_1, \xi_{K+p+1}]$ which makes it very simple 
to implement Dirichlet boundary conditions like the ones given in Eqn.~\eqref{eq:initial_final_condition}.

\begin{figure}[ht]
\centering
\includegraphics[width=0.45\textwidth]{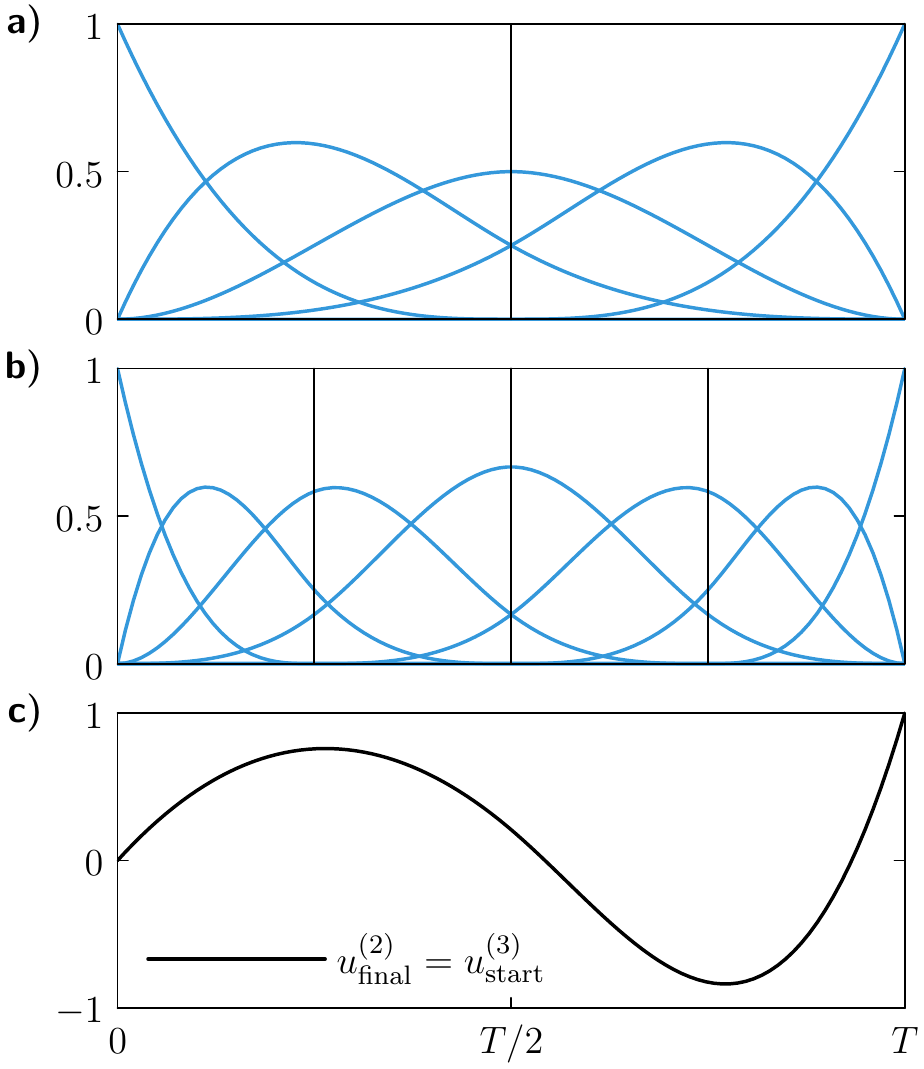}
\caption{
Illustration of the B-spline refinement process.
\textbf{\textsf{a})}~B-spline basis functions corresponding to the knot vector $\Xi_2$
defined in Eqn.~\eqref{eq:knot_vector_2}.
\textbf{\textsf{b})}~B-spline basis functions corresponding to the knot vector $\Xi_3$ 
defined in Eqn.~\eqref{eq:knot_vector_3}.
Dark vertical lines represent the interior knots of the knot vectors.
\textbf{\textsf{c})}
Representation of one and the same function
using the B-spline basis functions of the second and third B-spline level, respectively.
The final solution $u_\mathrm{final}^{(2)}$ of the optimal control problem corresponding 
to the second B-spline level serves as an initial guess $u_\mathrm{start}^{(3)}$ for 
the optimal control problem in the third B-spline level.
}
\label{fig:figure_refinement}
\end{figure}

The B-spline basis functions corresponding to the knot vectors $\Xi_2$ and $\Xi_3$ are shown in
Fig.~\ref{fig:figure_refinement}~a) and Fig.~\ref{fig:figure_refinement}~b), respectively.
Furthermore, Fig.~\ref{fig:figure_refinement}~c) shows a given control input using two representations
\begin{subequations}
\begin{align}
\label{eq:u_final_2}
u_\mathrm{final}^{(2)}(t) 
&= 
\sum_{k=1}^{K^{(2)}} 
c_k^{(2)} N_{k,p}^{(2)}(t),
\\
\label{eq:u_start_3}
u_\mathrm{start}^{(3)}(t) 
&= 
\sum_{k=1}^{K^{(3)}} c_k^{(3)} N_{k,p}^{(3)}(t)
\end{align}
\end{subequations}
corresponding to the basis functions of the second and third B-spline level, respectively.
These two representations of a single control input illustrate the basic idea behind 
the multilevel B-spline method.
Instead of solving the optimal control problem using the finest refinement level directly, it is 
often more promising to solve the optimization problem using a coarse representation of the control input first.
Subsequently the solution will be used as an initial guess for the optimal control problem in the 
next refinement level.
In Eqn.~\eqref{eq:u_final_2}, the solution of the optimization problem corresponding to the second B-spline level
is represented using $K^{(2)}$ coefficients $c_1^{(2)}, \ldots, c_{K^{(2)}}^{(2)}$.
The same function is represented in Eqn.~\eqref{eq:u_start_3} using $K^{(3)} > K^{(2)}$ 
coefficients corresponding to the knot vector $\Xi_3$.
However, the coefficients $c_1^{(3)}, \ldots, c_{K^{(3)}}^{(3)}$ do not represent an optimal solution 
with respect to the third B-spline level.
Rather, they serve as an excellent initial guess which will speed up the solution of the
next optimization problem significantly.

The solution corresponding to the previous B-spline level can be represented exactly by means 
of the new basis functions if the knots of the old knot vector are contained in the new knot vector.
An efficient numerical method to determine the coefficients in the new basis is called 
knot insertion \cite{prautzsch_2002}.
Alternatively, we may determine the coefficients of the next level 
$c_1^{(\ell+1)}, \ldots, c_{K^{(\ell+1)}}^{(\ell+1)}$
by solving a linear system corresponding to the equations
\begin{align*}
\sum_{k=1}^{K^{(\ell+1)}} c_k^{(\ell+1)} N_{k,p}^{(\ell+1)}(\hat{\xi}_m)
=
\sum_{k=1}^{K^{(\ell)}} c_k^{(\ell)} N_{k,p}^{(\ell)}(\hat{\xi}_m)
\end{align*}
for $m=1,\ldots,K^{(\ell+1)}$ with $\hat{\xi}_m$ being 
the  Greville points~\cite{johnson_2005, cottrell_2009} 
of the knot vector $\Xi_{\ell+1}$.


\section{Application to the dipolar droplet formation process}
\label{sec:application}

The numerical methods introduced in Section II, III and IV
will now be applied to optimize the self-bound dipolar droplet formation process.

\subsection{Parameterization, Constraints and Cost Function}

As mentioned in Section II, the control inputs are given by the 
s-wave scattering length $a_s$ and the trap frequencies 
$\omega_\rho$ and $\omega_z$.
While the system is initially a Bose-Einstein condensate, we are aiming to find optimal
trajectories of $a_s$, $\omega_\rho$ and $\omega_z$ which will smoothly transform this condensate 
into a single dipolar droplet.
The final droplet is stable in the sense that it will exist even 
after the external potential $V_\ext$ has been set to zero at final time $T$.

To this end, the physical control inputs
\begin{subequations}
\label{eq:parameterization_control_inputs}
\begin{align}
a_s(t) 
&= 
a_{s, \initial} \,+ (a_{s, \final}-a_{s, \initial}) \,\, u_1(t), \\
\omega_\rho(t) 
&= 
\omega_{\rho, \initial} + (\omega_{\rho, \final}-\omega_{\rho, \initial}) \, u_2(t), \\
\omega_z(t) 
&= 
\omega_{z, \initial} + (\omega_{z, \final}-\omega_{z, \initial}) \, u_3(t)
\end{align}
\end{subequations}
are parameterized using three individual 
B-splines $u_1(t)$, $u_2(t)$ and $u_3(t)$ with $t \in [0,T]$.
The initial and final values of $a_s$, $\omega_\rho$ and $\omega_z$ in Eqn.~\eqref{eq:initial_final_values} 
are taken into account using
\begin{subequations}
\label{eq:initial_final_conditions_b_spline}
\begin{align}
u_1(t=0) &= 0, \quad u_1(t=T) = 1, \\
u_2(t=0) &= 0, \quad u_2(t=T) = 1, \\
u_3(t=0) &= 0, \quad u_3(t=T) = 1.
\end{align}
\end{subequations}
As discussed at the end of Section II, we limit the allowed ranges of the control inputs
\[
a_s^\lb \leq a_s \leq a_s^\ub,
\;\;
\omega_\rho^\lb \leq \omega_\rho \leq \omega_\rho^\ub,
\;\;
\omega_z^\lb \leq \omega_z \leq \omega_z^\ub,
\] 
which can be easily translated to equivalent constraints for the B-spline functions
\begin{subequations}
\label{eq:constraints_b_spline_functions}
\begin{align}
u_1^\lb \leq u_1 \leq u_1^\ub,
\\
u_2^\lb \leq u_2 \leq u_2^\ub,
\\
u_3^\lb \leq u_3 \leq u_3^\ub.
\end{align} 
\end{subequations}
Each of these three B-spline functions are uniquely defined by the basis functions correponding 
to the knot vectors~\eqref{eq:knot_vectors} and the coefficients
\begin{align*}
\bm{c}_1^{(\ell)}
&= 
\big[
c_{1,1}^{(\ell)}, c_{1,2}^{(\ell)}, \ldots, c_{1,K_\ell-1}^{(\ell)}, c_{1,K_\ell}^{(\ell)} 
\big],
\\
\bm{c}_2^{(\ell)}
&= 
\big[
c_{2,1}^{(\ell)}, c_{2,2}^{(\ell)}, \ldots, c_{2,K_\ell-1}^{(\ell)}, c_{2,K_\ell}^{(\ell)} \big],
\\
\bm{c}_3^{(\ell)}
&= 
\big[
c_{3,1}^{(\ell)}, c_{3,2}^{(\ell)}, \ldots, c_{3,K_\ell-1}^{(\ell)}, c_{3,K_\ell}^{(\ell)} \big],
\end{align*}
where $\ell$ denotes the level of refinement and $K_\ell$ is the number of B-spline basis functions.
An important feature of B-spline functions is called the 
convex hull property~\cite{d_boor_1978}, which implies that we can implement
the constraints~\eqref{eq:constraints_b_spline_functions} by simply replacing
$u_1$, $u_2$ and $u_3$ with the corresponding coefficients
\begin{subequations}
\label{eq:constraints_c1_c2_c3}
\begin{align}
u_1^\lb \leq \bm{c}_1^{(\ell)} \leq u_1^\ub,
\\
u_2^\lb \leq \bm{c}_2^{(\ell)} \leq u_2^\ub,
\\
u_3^\lb \leq \bm{c}_3^{(\ell)} \leq u_3^\ub.
\end{align}
\end{subequations}

For a given knot vector $\Xi_\ell$ of dimension $s_\ell$ we have $K_\ell = s_\ell-(p+1)$ basis functions, 
see, e.g., Fig.~\ref{fig:figure_refinement}.
However, as a result of Eqn.~\eqref{eq:initial_final_conditions_b_spline}, 
the coefficients
\[
c_{1,1}^{(\ell)} = c_{2,1}^{(\ell)} = c_{3,1}^{(\ell)} = 0,
\quad
c_{1,K_\ell}^{(\ell)} = c_{2,K_\ell}^{(\ell)} = c_{3,K_\ell}^{(\ell)} = 1
\]
are already fixed.
Therefore, the unknown variables of the $\ell$th optimization problem are given by
\begin{equation}
\label{eq:coefficients_nlp}
\begin{aligned}
\bm{c}^{(\ell)}
=
\big[
c_{1,2}^{(\ell)}, \ldots, c_{1,K_\ell-1}^{(\ell)},
&c_{2,2}^{(\ell)}, \ldots, c_{2,K_\ell-1}^{(\ell)},\\
&\quad c_{3,2}^{(\ell)}, \ldots, c_{3,K_\ell-1}^{(\ell)}
\big].
\end{aligned}
\end{equation}
It can be easily verified that the total number of unknown coefficients
in the $\ell$th B-spline level is given by $6$, $9$, $15$ and $27$,
respectively.

For the cost function we chose
\begin{equation}
\label{eq:cost_function_droplet_problem}
\mathcal{J}(\psi(T)) = \big( \NParticles_0 - | \langle \psi_d, \psi(T) \rangle | \big)^2
\end{equation}
with $\psi(T)$ being implicitly dependent on the coefficients $\bc^{(\ell)}$. 
Furthermore, $\psi_d$ denotes the desired single droplet state.
In this context, it should be noted that the densities of the initial 
and desired states $\psi_0$ and $\psi_d$ are normalized to the same number of 
particles $\NParticles_0$.
Furthermore, the norm of $\psi(T)$ is strictly smaller than the norm
of $\psi_d$, which is a direct consequence of the
three-body loss term in the generalized GPE \eqref{eq:generalized_gpe}. 

The effect of the three-body loss term is noticeable during the whole time evolution.
Initially, the effect might be small as the density of the initial state is comparatively low. 
However, as soon as the wavefunction comes close to the target state $\psi_d$,
a significant number of atoms will inevitably get lost.
Regardless of how we select the control inputs, it is therefore impossible to make all atoms 
contribute to the desired droplet state at final time $T$.
Nonetheless, we will see that our optimization algorithm will find solutions minimizing the effect 
of the three-body loss term such that almost all particles
of the initial condensate will be transferred to the desired single droplet state.

The final optimal control problem in the $\ell$th level of refinement will be solved using 
an advanced implementation of the SQP method, which allows to take into account 
the box constraints~\eqref{eq:constraints_c1_c2_c3} in a highly efficient way.
Alternatively, the given constrained optimization problem could be reformulated 
as an unconstrained optimization problem.
To this end, the cost function \eqref{eq:cost_function_droplet_problem}
would have to be extended by additional penalty terms explicitly depending
on the coefficients~$\bm{c}^{(\ell)}$.

\subsection{Implementation Details}

The numerical methods presented above have been implemented in Matlab R2018a
on a powerful workstation computer.
The most time-consuming part in our optimization strategy is the numerical solution of the 
generalized Gross-Pitaevskii equation~\eqref{eq:generalized_gpe}.
Just like for the ordinary GPE \cite{mennemann_2015}, the computations can be accelerated significantly 
by means of a powerful graphics processing unit (GPU). 
More precisely, we employ a NVIDIA Quadro GV100 (32GB HBM2 GPU Memory) 	
being able to deliver more than $7.4$ TFLOPS of double precision 
floating-point (FP64) performance.

The numerical solution of the generalized GPE is based on the time-splitting spectral method (TSSM)
using a spectrally accurate algorithm to evaluate the dipolar interaction potential
\eqref{eq:dipolar_interaction} as outlined in Section \ref{sec:numerical_discretization}.
In order to take into account the elongated shape of the desired droplet state, we choose the
side lengths of the computational domain~\eqref{eq:Omega}
to be $L_x = 12\,\mu$m, $L_y = 12\,\mu$m and $L_z = 24\,\mu$m.
Our highly accurate implementation of the TSSM converges
at an exponential rate with respect to the number of spatial grid points 
$J_x$, $J_y$ and $J_z$.
In fact, we employ two sets of discretization parameters.
The first set is characterized by the spatial discretization parameters
$J_x = 72$, $J_y = 72$, $J_z = 64$ and a time step size of $\triangle t = 0.005$\,ms.
This rather coarse discretization will be used to solve the generalized GPE in the optimization
process.
As a cross check we will employ a second set of discretization parameters
given by $J_x = 2 \times 72$, $J_y = 2 \times 72$, $J_z = 2 \times 64$ and a time step size
of $\triangle t = 0.0025$\,ms.
In fact, the graphs of all physical quantities presented in the next section
correspond to numerical simulations using the second set of parameters.
However, the graphs are indistinguishable from the ones obtained with the set of coarse discretization parameters.

Our aim is to prepare a self-bound dipolar droplet state on a very short time scale
of $T=2$\,ms corresponding to one-tenth of the duration of the piecewise linear ramps
employed in \cite{baillie_2016}.
Accordingly, only $\NTimeSteps=400$ ($\NTimeSteps=800$) time steps are needed to compute the wave function $\psi(T)$
at final time $T$.
As a result of the small number of required spatial grid points $J=J_x J_y J_z$ 
and time steps~$\NTimeSteps$, the computational time to compute $\psi(T)$, 
and hence to evaluate \eqref{eq:cost_function_droplet_problem},
is less than four seconds in case of the coarse discretization and much shorter than
a minute in case of the fine discretization.
It seems that the potential of the above mentioned GPU is not fully exploited 
with the first set of discretization parameters. 
Therefore, the solution times corresponding to the second set of discretization parameters 
are significantly shorter than expected from a purely computational complexity 
based point of view.

Before starting our optimization algorithm we need to compute
the initial state $\psi_0$ and the desired state $\psi_d$.
As in the case of the ordinary GPE, these states can be computed using imaginary 
time propagation~\cite{chiofalo_2000}, i.e., the time step  $\triangle t$ 
in our implementation of the TSSM is replaced by $-i \triangle t$ and the density of 
the wave function is normalized to the number of desired atoms after every time step.

\subsection{Results}

In accordance with Eqn.~\eqref{eq:coefficients_nlp},
the initial point of the NLP corresponding to the first B-spline level
is given by a $6$-dimensional coefficient vector~$\bm{c}_0^{(1)}$.
The coefficients are chosen such that the corresponding control inputs $a_s$, $\omega_\rho$
and $\omega_z$ are simple linear ramps.

\begin{figure}[ht]
\centering
\includegraphics[width=0.45\textwidth]{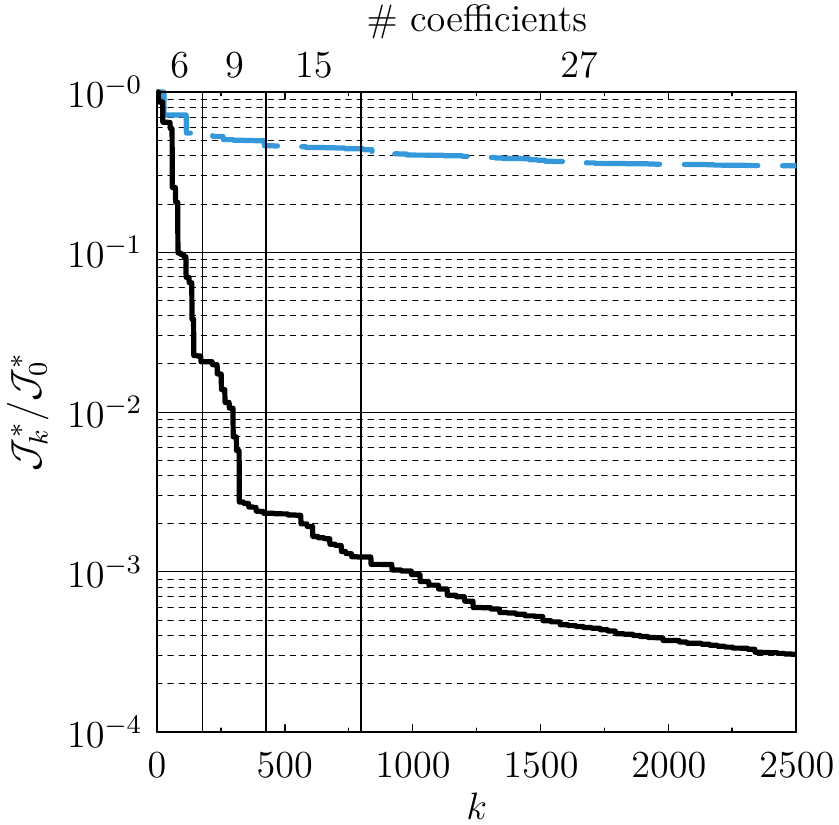}
\caption{
Normalized convergence history
corresponding to two different optimization strategies.
Solid black line: multilevel B-spline optimization using level $1$, $2$, $3$ and $4$.
Broken blue line: direct optimization using the finest B-spline 
parameterization (level $4$) only.
}
\label{fig:figure_convergence}
\end{figure}

\begin{figure*}[ht]
\centering
\includegraphics[width=1.0\textwidth]{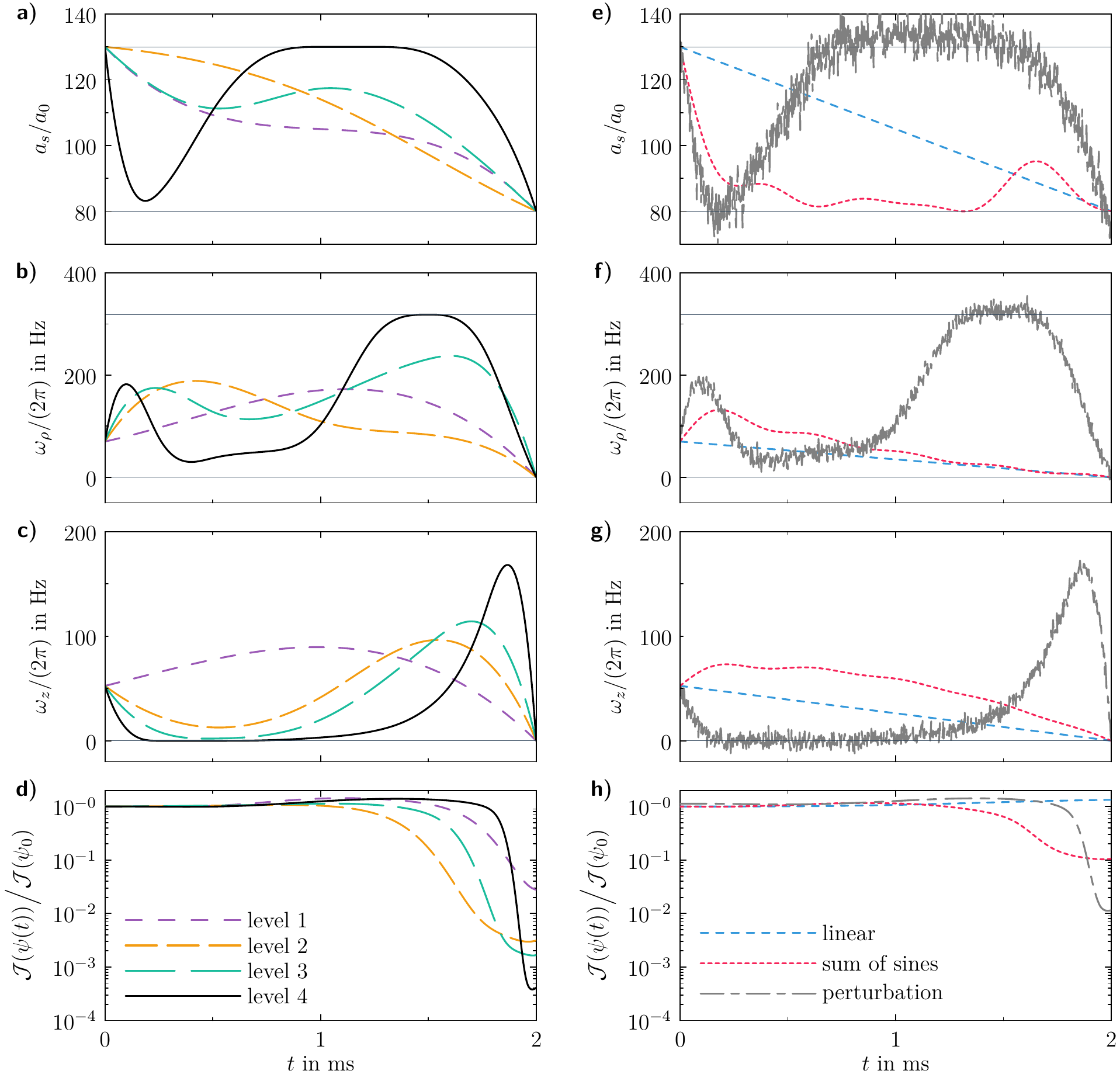}
\caption{
\textbf{\textsf{a})--\textsf{d})}
Time evolution of the control inputs $a_s$, $\omega_\rho$, $\omega_z$
and the corresponding (normalized) cost function
as a result of the multilevel B-spline optimization process.
Thin horizontal lines indicate the relevant lower and upper bounds of the control inputs.
\textbf{\textsf{e})--\textsf{h})}
Time evolution of another three realizations of the control inputs and their corresponding cost function.
}
\label{fig:figure_time_evolution_controls}
\end{figure*}

First, we study the convergence history of our optimization algorithm. 
To this end, we consider the best cost function value
\begin{equation*}
\mathcal{J}_k^* = \min_{r = 1, \ldots, k} \mathcal{J}_r
\end{equation*}
among the first $k$ cost function evaluations of the underlying SQP method.
The normalized convergence history is depicted in Fig.~\ref{fig:figure_convergence} for two different 
optimization strategies.
For normalization we employ the cost function value corresponding to the case where
all control inputs are given by simple linear ramps.

In the first strategy, we employ the original multilevel B-spline approach outlined
in Section~\ref{sec:multilevel_B_spline_method}.
In this context, it should be noted that we employ $15$ iterations of the SQP method
before triggering a new B-spline refinement operation.
However, the maximum number of iterations in the fourth B-spline level is set 
to infinity and the optimization is stopped after $2500$ cost function evaluations.
The corresponding convergence history is shown in Fig.~\ref{fig:figure_convergence}
as a black solid line.

Alternatively, we can start the optimization algorithm using the fourth level of refinement
right from the beginning.
The initial point of the SQP algorithm is given by a $27$-dimensional 
coefficient vector~$\bm{c}_0^{(4)}$.
As before, the coefficients are chosen such that the corresponding control inputs 
are simple linear ramps.
The resulting convergence history is depicted as a broken blue line in
Fig.~\ref{fig:figure_convergence}.

By comparing the results of both strategies it is clear that the multilevel
B-Spline approach is far superior to the approach in which the finest B-spline parameterization
is used right from the beginning. 
The successive refinement strategy prevents the optimization algorithm from getting stuck in one of 
the many very unfavorable local minima existing in a high-dimensional parameter space. 
In fact, the multilevel B-spline optimization method exhibits some similarities 
with multigrid preconditioning techniques frequently applied in numerical
linear algebra problems such as elliptic partial differential equations
\cite{hackbusch_2013}.

The time evolution of the control inputs $a_s$, $\omega_\rho$, $\omega_z$ 
and the corresponding (normalized) cost function
as a result of the multilevel B-spline optimization process is shown 
in Fig.~\ref{fig:figure_time_evolution_controls} a)--d).
It can be seen that $\mathcal{J}(\psi(T))$ 
decreases with increasing B-spline level and that the optimized control inputs are able to 
remain on their respective upper or lower bounds for time intervals of non-vanishing lengths. 
This particular behavior is a result of the compact support of the B-spline basis functions 
and could not be obtained by more simple control vector parameterizations like the 
sum of sines ansatz given in Eqn.~\eqref{eq:sum_of_sines} or Fourier type parameterizations in general.

As an example, we apply the parameterization in Eqn.~\eqref{eq:sum_of_sines} using $K=9$ 
sine functions for each of the three controls $a_s$, $\omega_\rho$ and $\omega_z$.
In contrast to the multilevel B-spline method, the box constraints on the control inputs
cannot be easily transferred to the corresponding coefficients $c_k$, $k=1, \ldots, K$. 
We therefore resort to quadratic penalty functions sanctioning the squares of the constraint violations \cite{nocedal_wright_2006}.
Since the SQP algorithm does not perform well for the given unconstrained optimization problem,
we apply the Nelder-Mead algorithm using $2500$ cost function evaluations as a stopping criterion.
Furthermore, the initial guess $\bm{c}_0 \in \R^{27}$ corresponds to linear ramps, meaning that all coefficients 
of the initial control vector are set to zero.
The numerical results of this alternative optimization strategy are depicted in 
Fig.~\ref{fig:figure_time_evolution_controls} e)--h).
As can be seen from Fig.~\ref{fig:figure_time_evolution_controls} e), the lower bound of the 
control input $a_s$ becomes active only at one or two isolated points in time.
Moreover, Fig.~\ref{fig:figure_time_evolution_controls} h) shows that the corresponding 
final value of the objective function is almost two orders of magnitude larger than the final value
corresponding to the multilevel B-spline method.

The above reported numerical results turned out to be stable with respect to variations of the initial
guess used to initialize the sum of sines and multilevel B-spline optimization algorithms.
More precisely, having tested a dozen of randomly selected initial conditions we always found that the final cost function value of the multilevel B-spline method was about two orders of magnitude smaller than the corresponding value of the sum of sines approach.

No reduction of the cost function value $\mathcal{J}(\psi(T))$ can be observed in the case where the 
control inputs are given by simple linear ramps, see Fig.~\ref{fig:figure_time_evolution_controls} h).
In fact, much larger final times $T$ are needed to observe a significant improvement.
However, with increasing $T$
more and more particles are lost as a result of the three-body loss term in the
generalized Gross-Pitaevskii equation~\eqref{eq:generalized_gpe}.
For that reason, and in strong contrast to most quantum control applications, 
the final desired state cannot be reached by adiabatically changing the control inputs.
Finally, we would like to stress that the three-body loss term is of crucial importance for a realistic 
description of dipolar Bose-Einstein condensates and hence cannot be omitted.
This is true in particular if the final state comes close to the desired state which is characterized by a
much higher peak density than the initial state.
In this context it is important to remember that the three-body loss term in the generalized GPE~\eqref{eq:generalized_gpe} is proportional to $|\psi|^4$.

\begin{figure*}[ht]
\centering
\includegraphics[width=1.0\textwidth]{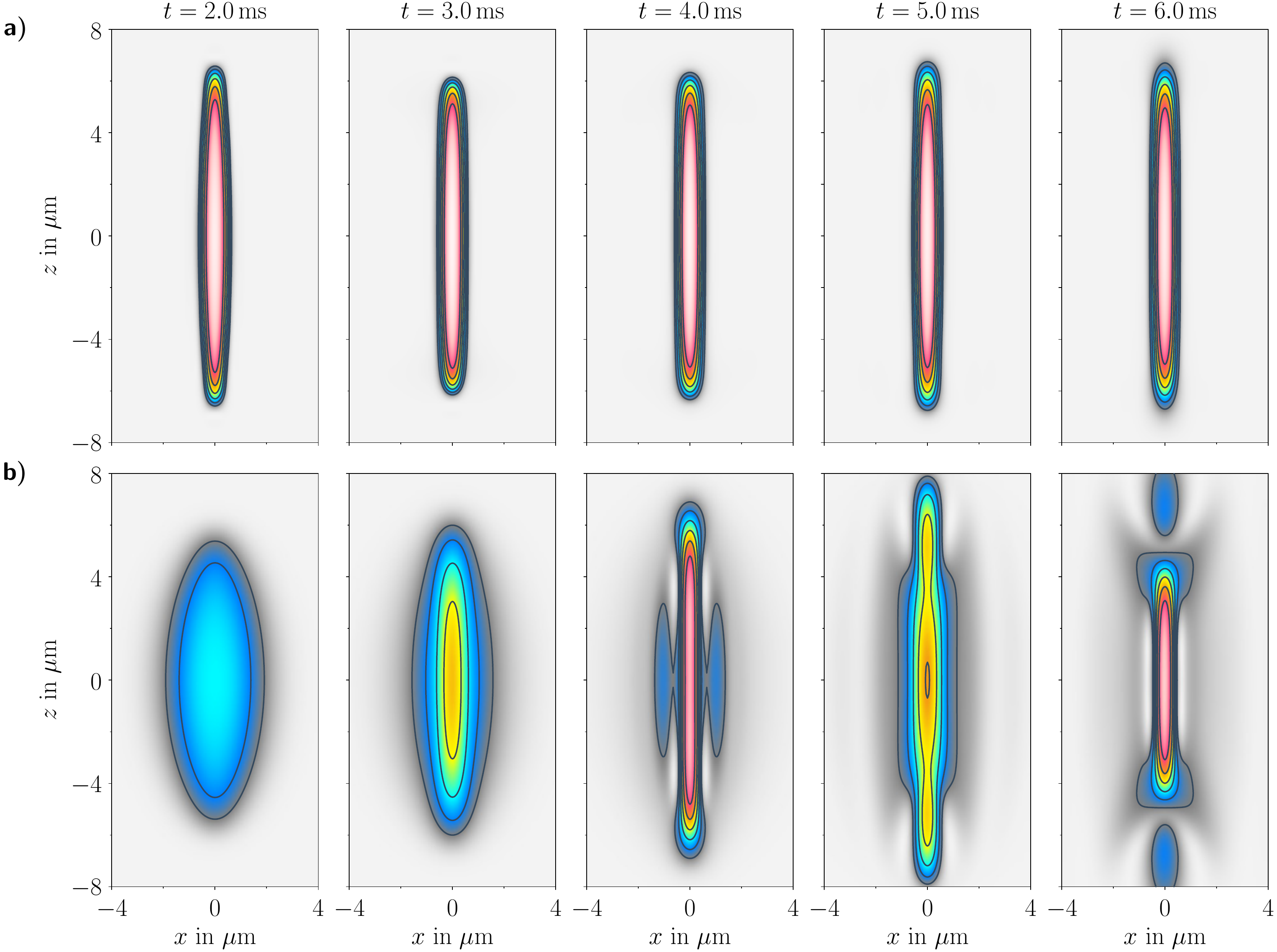}
\caption{
Time evolution
of the density at $y\equiv 0$ using the optimized controls \textbf{\textsf{a})} and
the linear controls~\textbf{\textsf{b})} at various times after the final time $T$.
The actual computational domain is given by
$\Omega = [-6\,\mu\mathrm{m},6\,\mu\mathrm{m}] \times [-6\,\mu\mathrm{m},6\,\mu\mathrm{m}] \times
[-12\,\mu\mathrm{m},12\,\mu\mathrm{m}]$.
}
\label{fig:figure_snapshots_xz}
\end{figure*}

\begin{figure*}[ht]
\includegraphics[width=1.0\textwidth]{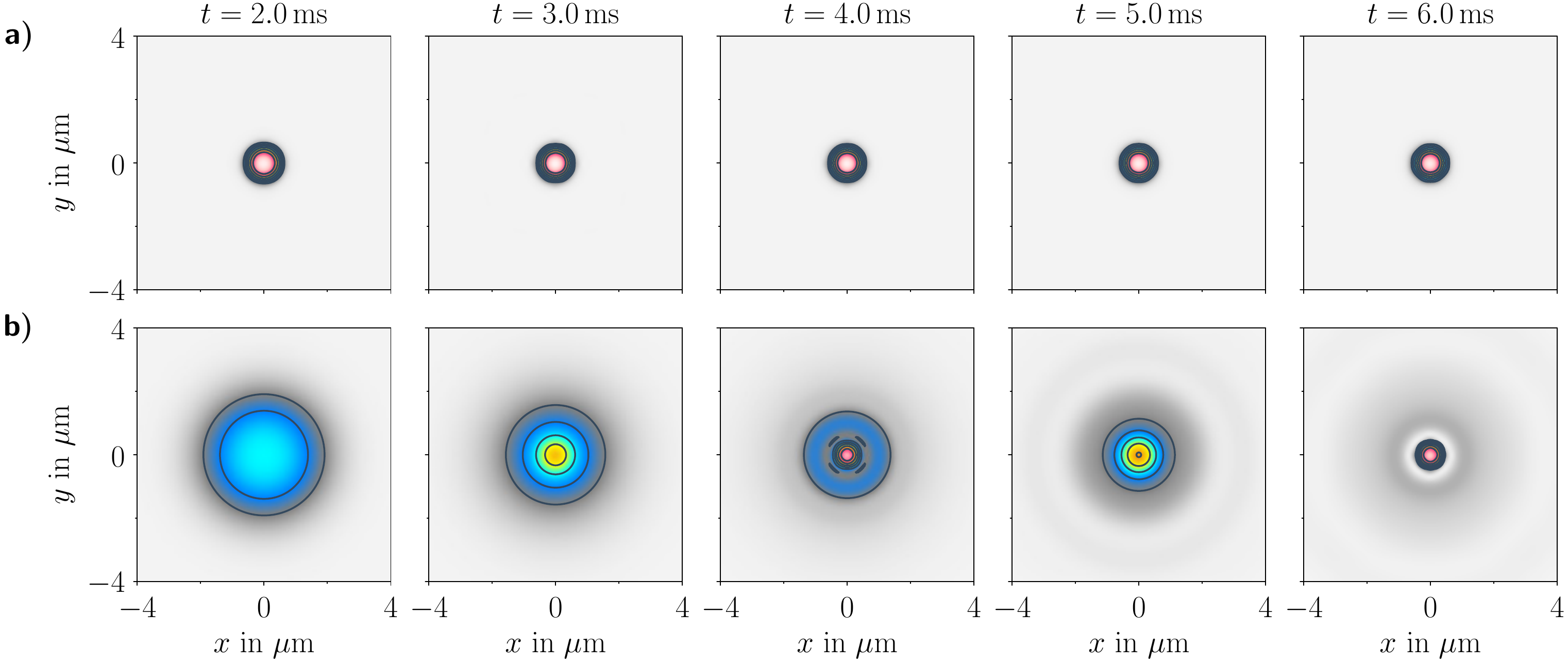}
\caption{
Time evolution of the density at $z\equiv 0$ using the optimized controls \textbf{\textsf{a})} and
the linear controls~\textbf{\textsf{b})} at various times after the final time $T$.
The actual computational domain is given by
$\Omega = [-6\,\mu\mathrm{m},6\,\mu\mathrm{m}] \times [-6\,\mu\mathrm{m},6\,\mu\mathrm{m}] \times
[-12\,\mu\mathrm{m},12\,\mu\mathrm{m}]$.
}
\label{fig:figure_snapshots_xy}
\end{figure*}

Several snapshots of the density at $y\equiv 0$ and $z\equiv 0$ are depicted in 
Fig.~\ref{fig:figure_snapshots_xz} and Fig.~\ref{fig:figure_snapshots_xy}, respectively.
They illustrate the time evolution of the density at various times after the final time $T$.
In the simulations 
we set
$a_s(t) = a_{s, \final}$,
$\omega_\rho(t) = \omega_{\rho, \final}$
and $\omega_z(t) = \omega_{z, \final}$ 
for $t \geq T$.
The density corresponding to the optimized control inputs can be clearly identified
as the density of a stable self-bound dipolar droplet state.
In contrast, shape and amplitude of the density corresponding to the linearly decreasing 
control inputs show very strong oscillations.

\begin{figure*}[ht]
\centering
\includegraphics[width=1.0\textwidth]{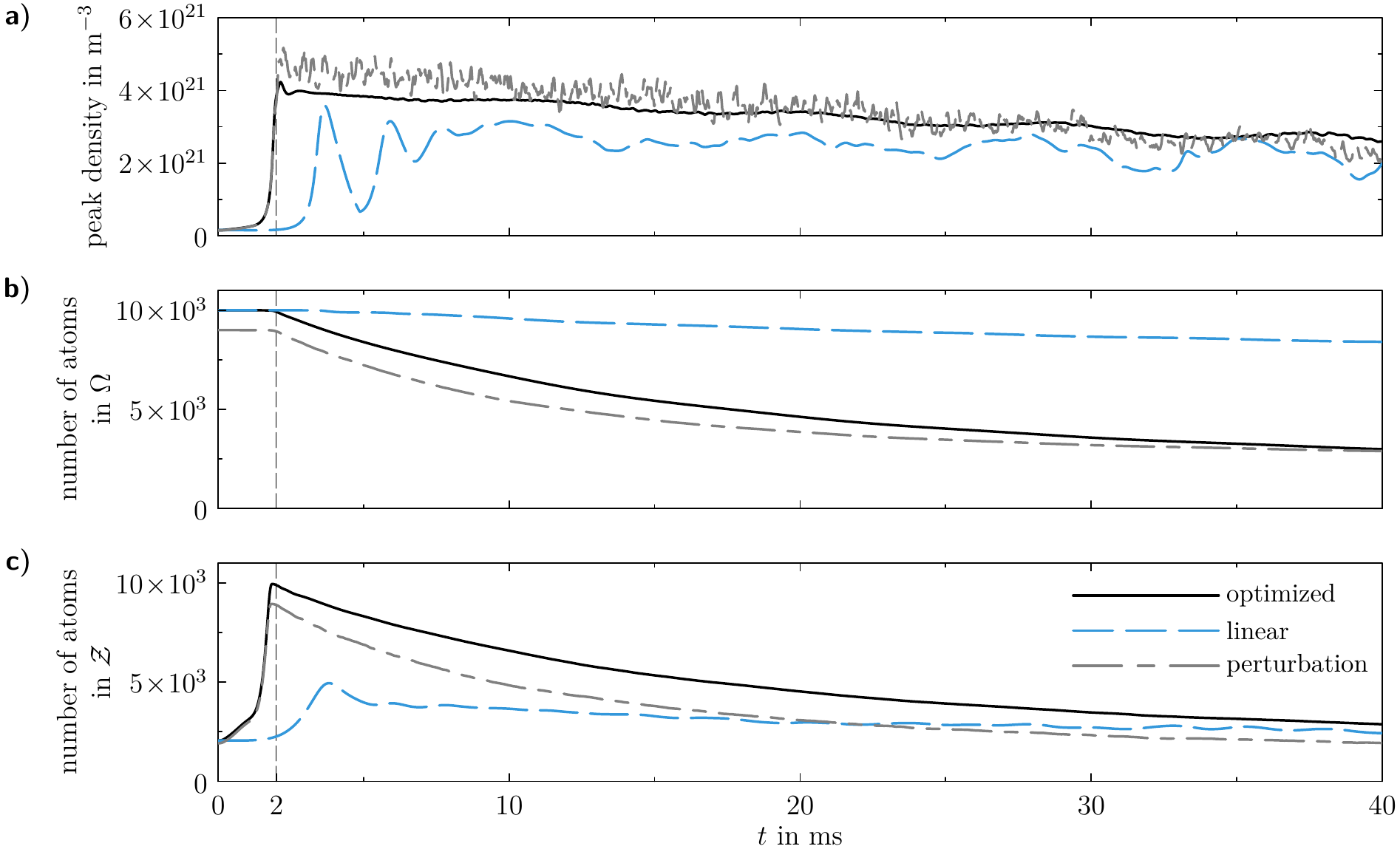}
\caption{
Time evolution of several observables for three different simulation scenarios.
Black solid line: numerical results corresponding to the optimized
control inputs using the multilevel B-spline method.
Blue broken line: linearly decreasing control inputs.
Grey broken line: optimized control inputs including additive white Gaussian noise and a systematic 
perturbation over the whole time interval $[0,T]$.
Furthermore, the number of atoms of the initial state has been reduced to $\tilde{\mathcal{N}}_0=9000$ particles.
\textbf{\textsf{a})} Time evolution of the peak density.
\textbf{\textsf{b})} Time evolution of the number of atoms 
in the computational domain $\Omega$.
\textbf{\textsf{c})} Time evolution of the number of atoms 
in a cylindrical volume 
$\mathcal{Z}$ closely surrounding the density of the desired single droplet state.
}
\label{fig:figure_time_evolution_observables}
\end{figure*}

These oscillations become even more obvious if we consider the time evolution of the peak 
density as shown in Fig.~\ref{fig:figure_time_evolution_observables} a). 
Moreover, it is interesting to note, that the optimization algorithm finds control inputs
which keep the peak density low until shortly before the end of the optimization at $T=2$\,ms. 
In this way the effect of the three-body loss term is minimized and less than $100$ atoms are 
lost during the preparation of the self-bound dipolar droplet state. 
This corresponds to more than $99\%$ of the initial atoms contributing to the desired final state. 
However, due to the high density of the final state, the effect of the three-body loss 
term becomes even more relevant for $t>T$ as can be seen from 
Fig.~\ref{fig:figure_time_evolution_observables}~b).
In case of the linear ramps, the effect of the three-body loss term is rather small, which
can be explained by the comparatively small peak (and average) density
which deviates strongly from the density of the target state $\psi_d$.

To investigate the above mentioned effect in more detail we take a look at
Fig.~\ref{fig:figure_time_evolution_observables}~c)
showing the time evolution of the number of atoms 
in the cylindrical domain
\begin{equation}
\label{eq:C}
\begin{aligned}
\mathcal{Z}
=
\big\{
(x,y,z) \in \Omega \; &\big| \; \sqrt{x^2 + y^2} \leq 0.75\,\mu \mathrm{m},
\;
\\
&\qquad |z| \leq 7.5\,\mu \mathrm{m}
\big\}.
\end{aligned}
\end{equation}
The parameters in~\eqref{eq:C} are chosen such that
the density of the desired droplet state is closely surrounded.
As expected, almost all atoms of the initial state are transferred to $\mathcal{Z}$ 
for the case of the optimized control inputs. 
However, in case of the linear ramps, strong fluctuations of the density cause 
the atoms to get distributed 
over the whole computational domain $\Omega$.
While the overall loss of atoms as a result of the three-body loss term is small,
a comparatively small fraction of these atoms actually contributes 
to the desired self-bound droplet state.

With regard to real-world quantum experiments it is important to characterize 
the stability of the control trajectories found by the optimization algorithm.
In this context, we consider the optimized B-spline functions $u_1$, $u_2$ und $u_3$
in combination with small perturbations of the initial and final values 
\begin{equation*}
\begin{aligned}
\tilde{a}_{s, \initial} &= 1.03 \, a_{s,\initial},
\\
\tilde{a}_{s, \final}   &= 0.97 \, a_{s,\final},
\\
\tilde{\omega}_{\rho, \initial} &= 1.03 \,\omega_{\rho, \initial},
\\
\tilde{\omega}_{z, \initial} &= 0.97 \,\omega_{z, \initial}
\end{aligned}
\end{equation*}
in Eqn.~\eqref{eq:parameterization_control_inputs}.
The result is a systematic perturbation of the control inputs over the whole time interval $[0,T]$.
With these modifications we model perturbations typically occurring in experiments, 
be it uncertainties in the interaction strength or in the atom number.
Additionally, we add white Gaussian noise to the control inputs
to model fluctuations, e.g. in the magnetic fields to control the scattering length.
The variance of the individual noise signals is given by $\sigma^2$ with a standard deviation of $\sigma=0.03$.
Moreover, the noise signals are scaled with the maximum values of the control inputs, 
see Fig.~\ref{fig:figure_time_evolution_controls} e)--g).
However, the main instability in BEC experiments is the number of particles $\mathcal{N}_0$ 
trapped in the potential at $t=0$.
This effect is taken into account by artificially lowering $\mathcal{N}_0$ 
by $10$ percent, which is a typical value reached in experiments.
In other words, the initial state of the perturbed simulation is recalculated 
using $\tilde{\mathcal{N}}_0=9000$ particles, whereas
the computations of the multilevel B-spline method where based on the assumption that initially 
$\mathcal{N}_0=10000$ particles are trapped in the quadratic potential at $t=0$.
Despite these relatively strong perturbations, we still observe a remarkable reduction of the objective function value
at final time $T$, 
see Fig.~\ref{fig:figure_time_evolution_controls} h). 
The time evolution of the corresponding observables
is shown in Fig.~\ref{fig:figure_time_evolution_observables}
along with the results of the previous simulations.
While the effect of the perturbations is clearly visible, the outcome of the 
numerical experiment remains largely the same.

\section{Conclusion and outlook}

We have introduced advanced numerical methods for the optimization of the self-bound dipolar 
droplet formation process.
By means of a multilevel B-spline control vector parameterization we were able to dramatically speed up
the transfer of a given Bose-Einstein condensate to this exotic quantum object.
Moreover, it has been seen that the optimized control inputs minimize the effect of the three-body loss term
in the generalized GPE and hence almost all atoms of the initial state could be transferred 
to the desired state at final time $T$.
In order to reduce the overall numerical costs we have also shown how to evaluate the dipolar interaction potential
using recently introduced spectrally accurate algorithms.
Our considerations are not limited to the formation of a single self-bound dipolar droplet state.
Rather, we are convinced that the presented algorithms can directly be applied to many other 
interesting parameter regimes, e.g. to efficiently realize the elusive supersolid state of matter~\cite{wenzel_2017,tanzi_2018}.

\section{Acknowledgement}
We thank Matthias Wenzel for discussions. 
This project has received funding from the European Union’s Horizon 2020 research and 
innovation programme under the Marie Sk\l odowska-Curie grant agreement No. 746525. 
TL acknowledges financial support by the Baden-Württemberg Stiftung through the Eliteprogramme 
for Postdocs. 
LE acknowledges support from the Austrian Science Foundation (FWF) under grant No. P31140-N32 (project 'ROAM').
JFM acknowledges support from the Austrian Science Foundation (FWF) under grant No. 
P32033, (“Numerik für Vielteilchenphysik und \textit{Single-Shot} Bilder”).
NJM acknowledges support from the Austrian Science Foundation (FWF) 
under grant No. F65 (SFB “Complexity in PDEs”) and grant No. W1245 (DK "Nonlinear PDEs").
JFM and NJM acknowledge the Wiener Wissenschafts- und TechnologieFonds (WWTF) 
project No. MA16-066 (“SEQUEX”).

\bibliographystyle{elsarticle-num}
\bibliography{bec}


\end{document}